\documentclass[prb,twocolumn,showpacs,showkeys,preprintnumbers,superscriptaddress,amsmath,amssymb]{revtex4}
\usepackage{graphicx}
\usepackage{dcolumn}
\usepackage{bm}
\usepackage[latin1]{inputenc}
\usepackage{psfrag}
\usepackage{epsfig}
\long\def\comment#1{}

\begin{document}
\title{\textbf{\large RVB gauge theory and the Topological  degeneracy 
in the Honeycomb Kitaev model}}

\author{Saptarshi Mandal}
\affiliation{Theoretical Physics Division, IACS, Jadavpur, Kolkata-700032. India.}
\author{R. Shankar}
\affiliation{The Institute of Mathematical Science, CIT Campus, Chennai-600113, India.} 
\author{ G. Baskaran}
\affiliation{The Institute of Mathematical Science, CIT Campus, Chennai-600113, India.} 
\affiliation{Perimeter Institute for Theoretical Physics,
31 Caroline St N, Waterloo, Ontario, Canada N2L 2Y5}
\begin{abstract}

We relate the Z$_2$ gauge theory formalism of the Kitaev model to the $SU(2)$
gauge theory of the resonating valence bond (RVB) physics. Further, we
reformulate a known \cite{jw1,jw2,jw3,jw4} Jordan-Wigner transformation of
Kitaev model on a torus in a general way that shows that it can be thought of
as a Z$_2$ gauge fixing procedure. The conserved quantities simplify in terms
of the gauge invariant Jordan-Wigner fermions, enabling us to construct exact
eigen states and calculate physical quantities. We calculate the fermionic
spectrum for flux free sector for different gauge field configurations and
show that the ground state is four-fold degenerate on a torus in thermodynamic
limit. Further on a torus we construct four mutually anti-commuting operators
which  enable us to prove that all eigenstates of this model are four
fold degenerate in thermodynamic limit. 

\end{abstract}

\maketitle
\section{Introduction}

Solid state realization of q-bits that do not decohere easily is a challenging
task in the field of Quantum Computation. Topological defects in strongly
correlated quantum many body systems are protected from decoherence and have
been suggested as q-bit candidates \cite{qcgen1,qcgen2,qcgen3,qcgen4,qcgen5}.
In this context, Kitaev constructed a remarkable two dimensional quantum spin
model that exhibits abelian and non-abelian anyons and is exactly solvable for
its spectrum \cite{kit1,kit2}. It was later shown that the spin correlation
functions are also exactly solvable \cite{short}.  This model is also extremely
interesting form point of view of frustrated spin models and the physics of the
resonating valence bond (RVB)
states\cite{rvb-dm1,rvb-dm2,rvb-dm3,rvb-dm4,rvb-dm5}. It realizes, in an exact
fashion, the phenomenon of quantum  number fractionization and emergent gauge
fields that were conjectured and approximately realized in 2D models for RVB
states or quantum spin liquids\cite{frac1,frac2}. It has also been shown that
the Jordan-Wigner transformation in this model yields a local fermionic
theory\cite{jw1,jw2,jw3,jw4}. This  makes the Kitaev model an important one
that warrants further investigation; and no wonder that an extensive body of
research \cite{othkit1,othkit2,othkit3,othkit4,othkit5} has already been
carried out  exploring its many fascinating aspects. Kitaev showed that the
model has a natural formulation in terms of a Majorana fermion interacting with
Z$_2$ gauge fields. The remarkable feature of the Kitaev model is that the
gauge fields turn out to be static. This greatly simplifies the dynamics
leading to the exact computation of the spectrum and spin-spin correlation
functions. 

The gauge theory of spin-$\frac{1}{2}$ models  has a recent history. It was
initially formulated \cite{bza} in the context of strongly correlated
electronic systems such as a spin-$\frac{1}{2}$ Mott insulator, as a way of
implementing the single electron occupancy constraint,
\begin{equation}
\label{u1gl}
\sum_{\sigma}\left(c^\dagger_{i\sigma}c_{i\sigma}-1\right)\vert\psi\rangle=0.
\end{equation}
This equation is the Gauss law constraint for a $U(1)$ lattice gauge theory.
It resulted in a strongly interacting U(1) gauge theory formalism of
spin-$\frac{1}{2}$ models. Soon it was realized that in spin-$\frac{1}{2}$
systems, the $U(1)$ gauge  invariance always implied an $SU(2)$ invariance
\cite{su21,su22,su23}.  The consequent $SU(2)$ gauge theory formalism was found
to be useful in the context of relating apparently different mean field
solutions of the model. The extended Hilbert space of the spin-$\frac{1}{2}$
system is much smaller than that of normal lattice gauge theory where the gauge
degrees of freedom on the links are $SU(2)$ group elements. Consequently, it
was shown that the essential physics of the spin-$\frac{1}{2}$ system is
captured by a Z$_2$ gauge theory \cite{bt,martson}, where the Z$_2$ gauge group is the
center of the original $SU(2)$ gauge group. The Z$_2$ gauge theory formalism
has been effectively used to bring out the physics of quantum number
fractionization in spin-$\frac{1}{2}$ systems \cite{frac2}.

In this paper we follow the route charted out above in the context of the
Kitaev model and show that the Z$_2$ gauge theory can indeed be thought of as
the center of the $SU(2)$ gauge theory of RVB theory. This sheds light on
Kitaev's assertion \cite{kit1} that the model represents the same universality
class of topological order as RVB. We examine the degeneracy of states in the 
system defined with periodic boundary conditions in detail. We show that
this degeneracy, which characterises the topological order in the system, 
arises from the so called large gauge transformations. Namely, gauge field 
configurations which correspond to the same flux configuration that are not
related to each other by local gauge transformations. These topologically
distinct gauge field configurations can be labelled by the value of the Wilson
loops that wind around the torus in the two different directions. We 
find these gauge configurations for the ground state and demonstrate the 
four-fold degeneracy of the ground state by explicity computing the energies.
Further, we generalise this proof to all eigenstates by constructing the
operators that generate the large gauge transformations and showing that
they do not change the energy of the system in the thermodynamic limit.

The rest of the paper is organised as folows. Section (\ref{km}) reviews the
Kitaev honeycomb model with its main features and some mathematical notions
which are used in later sections. The $SU(2)$ gauge symmetry of the model is
reviewed in section (\ref{su22z2}) where we show that Kitaev's choice of the
representation of the spin operators in terms of Majorana fermions amounts to a
$SU(2)$ gauge fixing procedure which fixes the gauge upto the center Z$_2$
gauge transformations.  In section (\ref{z2gtkm}), we construct a generalized
Jordan-Wigner transformation and show that it is a Z$_2$ gauge fixing
transformation in the Kitaev formalism. We derive the fermionic Hamiltonian and
simplify the conserved quantities in terms of gauge invariant Jordan-Wigner
operators. In section (\ref{gsd}), we explicitly work out the four fold
degeneracies of the Kitaev model on a torus. We derive the fermionic spectrum
for flux free sector in (\ref{gsd1})  and show that it leads to four degenerate
ground states in thermodynamic limit. Following this in section (\ref{tpg}), we
demonstrate that every eigenstate of the Kitaev model has four fold degeneracy.
To this end we derive  four mutually anticommuting operators that commute with
the hamiltonian in the thermodynamic limit. The minimum dimension of the
representation of the 4-dimensional Clifford algebra is 4. Thus we are able to
prove that all eigenstates are at least four fold degenerate.  We summarize our
results in the section (\ref{dis}).\\

\section{The Kitaev model}
\label{km}
\subsection{The Hamiltonian}
\label{modeldef}
\begin{figure}[h!]
\center{\hbox{\epsfig{figure=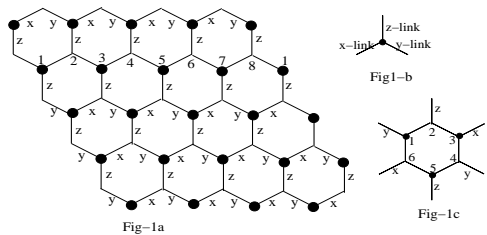,height=1.2in,width=2.5in}}}
\caption{Honeycomb lattice and the Kitaev model}
\label{kmd}
\end{figure}

The Kitaev model is a spin-$\frac{1}{2}$ system on a honeycomb
lattice. The Hamiltonian is
\begin{equation}
\label{1}
H=-J_{x}\sum_{{\langle ij\rangle}_x}\sigma^{x}_{i}\sigma^{x}_{j}
-J_{y}\sum_{{\langle ij\rangle}_y}\sigma^{y}_{i}\sigma^{y}_{j}
-J_{z}\sum_{{\langle ij\rangle}_z}\sigma^{z}_{i}\sigma^{z}_{j}, 
\end{equation}
where $i,j$ run over the sites of the honeycomb lattice,
$\langle ij\rangle_a,~a=x,y,z$ denotes the nearest 
neighbor links oriented in the $a$'th direction as shown in
Fig. \ref{kmd}. We will be working with periodic boundary conditions which
are defined as follows. The honeycomb lattice is a triangular lattice 
with a basis of two sites. The sites of the triangular lattice are given by,
\begin{equation}
\label{tlsites}
{\bf R}_{m,n}=m{\bf e}_1+n{\bf e}_2,
\end{equation}
where $m,~n$ are integers and ${\bf e}_1\cdot{\bf e}_2=-\frac{1}{2},~
{\bf e}_1\cdot{\bf e}_1=1={\bf e}_2\cdot{\bf e}_2$. The label $i$ of the sites
of the honeycomb lattice therefore stands for $(m,n,\alpha)$ where $\alpha=a,b$ is
the sub lattice label. The periodic boundary conditions are then defined by,
\begin{equation}
\sigma^a_{m,n,\alpha}=\sigma^a_{m+M,n+N,\alpha}.
\end{equation}
\subsection{The conserved quantities}
\label{spconsq}

There is a conserved quantity associated with every plaquette of 
the lattice. If the plaquette is denoted as $p$ and its vertices
labelled as shown in Fig. \ref{kmd}, then, following Kitaev's notation,
the conserved quantity is,
\begin{equation}
\label{bpdef}
B_p=\sigma_1^x\sigma_2^z\sigma_3^y\sigma_4^x\sigma_5^z\sigma_6^y.
\end{equation}
We have $B_p^2=1$ implying that the $B_p$s can take values $\pm 1$.
It is clear that any product of the $B_p$s will also commute with the
Hamiltonian. In fact there is a conserved quantity associated with every
closed self avoiding loop, $C$, on the lattice defined the following way.
At every site, the path will pass through two of the
three bonds that emanate from it. We call these two bonds as the tangential
bonds and the third one the normal bond. We associate two tangential vectors
at each site, ${\bf{\hat t}}_{1i}$ and ${\bf{\hat t}}_{2i}$
which are either ${\bf{\hat x}}$, ${\bf{\hat y}}$ or ${\bf{\hat z}}$
according to the direction of the incoming bond and the outgoing bond
respectively. We then define a normal vector ${\bf{\hat n}}_i$ as
\begin{equation}
\label{nidef}
{\bf{\hat n}}_i\equiv{\bf{\hat t}}_{1i}\times{\bf{\hat t}}_{2i}.
\end{equation}

If the sites of $C$ are $i_1,i_2,.....i_N$, then the 
conserved quantity associated with it is,
\begin{equation}
\label{bcdef}
B_C=\prod_{n=1}^N\left({\bf{\hat n}}_{i_n}\cdot{\bf\sigma}_{i_n}\right).
\end{equation}
It can be checked that,
\begin{equation}
[B_C,H]=0,~~~~B_C^2=1.
\end{equation} 

We will call $C$ topologically trivial if it can be written as a product 
of $B_p$s. On the torus, we have two loops which wind the torus around 
in the two directions which cannot be expressed as a product of $B_p$s. 
One cannot be obtained from the other by multiplication by $B_p$s. We will 
call these two (Wilson)loops $W_1$ and $W_2$. 

All the $B_p$s are not independent due to the identity,
\begin{equation}
\label{bpid}
\prod_p B_p=1.
\end{equation}
Thus there are $N_p-1$ independent $B_p$s, where $N_p=MN$ is the number
of plaquettes. Together with $W_1$ and $W_2$, we have a total
of $N_p+1$ conserved quantities on the torus. These two loop operators
account for the four fold degeneracy on a torus.
\section{From $SU(2)$ to Z$_2$}
\label{su22z2}
\subsection{$U(1)$ and $SU(2)$ gauge symmetry}

Interacting quantum spin systems often lead to spontaneously broken symmetric
states such as a ferro, antiferro or spiral magnetic states. Low energy physics of these ordered states are captured by the well known spin wave approximations, reasulting in Goldstone mode type bosonic low energy effective theories \cite{fradkin}. A quantum spin liquid, on the other hand, has no classical long range order. Experiments in LaCuO$_{4}$ and other low spin Mott insulators, according to Anderson, indicated possible presence of neutral fermionic excitations in a quantum spin liquid \cite{nspin1,nspin2,nspin3}. A theory to describe such a quantum spin liquid or resonating valence bond (RVB) state needed a paradigm shift from spin wave theory. It was also clear that a quantum spin liquid, in view of different possible phase coherence among disordered spin configurations, could offer a variety of quantum spin liquid states to be realized in nature. RVB gauge theory attempted to capture these new possibilties, through an approach involving enlarged Hilbert space and emergent gauge fields in strongly correlated electron systems. Through the work of Wen\cite{WenPSG} and others it has become clear that there is a plethora of spin liquid phases, characterized by \textit{quantum order and projective symmetry groups.}

The spin-$\frac{1}{2}$ Hilbert space can be realized as the subspace
of the Hilbert space of two fermions defined by the constraint,
\begin{equation}
\label{u1cons}
\left(\sum_{\sigma=\uparrow\downarrow}~c^\dagger_\sigma c_\sigma -1\right)
\vert\psi\rangle=0,
\end{equation}
where $c^\dagger_\sigma$ and $c_\sigma$ are the fermion creation and
annihilation operators. As mentioned earlier, Eq.~(\ref{u1cons}) can
be looked upon as the Gauss law constraint for a $U(1)$ gauge theory. The
LHS of the equation being the generator of the following $U(1)$ gauge 
transformations on the fermion operators,
\begin{equation}
\label{u1trans}
c_\sigma\rightarrow e^{i\Omega}c_\sigma,~~~
c^\dagger_\sigma\rightarrow e^{-i\Omega}c^\dagger_\sigma.
\end{equation}
The spin operators,
\begin{equation}
\label{spindef}
S^a=\frac{1}{2}
c^\dagger_\sigma\sigma^a_{\sigma\sigma^\prime}c_{\sigma^\prime},
\end{equation}
are then the gauge invariant observables of the theory. $\sigma^a$ are the
Pauli spin matrices. 

The single occupancy constraint in the spin-$\frac{1}{2}$ theory implies
that a spin-$\uparrow$ hole is the same as a spin-$\downarrow$ particle in
the physical space. This can be mathematically expressed as an $SU(2)$
gauge invariance. It is convenient to express this symmetry in terms of 
a matrix of the fermion operators
\begin{equation}
\label{smatdef}
{ \Psi}\equiv\left(\begin{array}{cc}
c_\uparrow& -c^\dagger_\downarrow\\
c_\downarrow& c^\dagger_\uparrow
\end{array}\right).
\end{equation}
In terms of this matrix, the spin operators are given by,
\begin{equation}
\label{gtgs}
 S^a=\frac{1}{4}{\rm tr}~{ \Psi}^\dagger \sigma^a { \Psi}.
\end{equation}
The generators of the $SU(2)$ gauge transformation are given by,
\begin{equation}
\label{sps}
\tilde S^a=-\frac{1}{4}{\rm tr}~{ \Psi}\sigma^a{ \Psi}^\dagger.
\end{equation}
The ${ \Psi}$ matrix transforms under the $SU(2)$ spin and $SU(2)$ 
gauge transformations as,
\begin{equation}
\label{smtrans}
{ \Psi}\rightarrow U_S{ \Psi}U^\dagger_G.
\end{equation}
Where $U_S$ and $U_G$ are $SU(2)$ matrices representing the spin and 
gauge transformations respectively. It is clear from equations~ 
 (\ref{gtgs}), (\ref{sps}) and (\ref{smtrans}) that the spin operators are 
gauge invariant and the generators of gauge transformations are spin singlet.
The constraint in Eq.~(\ref{u1cons}) is exactly equivalent to the $SU(2)$
Gauss law,
\begin{equation}
\label{su2gl}
\tilde S^a\vert\psi\rangle=0.
\end{equation}

Before we close this section, we wish to mention that the above gauge theory formalism offers a possible way to understand quantum spin liquid states as and when they exist. This formalism does not gaurantee a simple gauge theory structure at all energy scales in the physics of the problem. It only suggests that in some systems (for some Hamiltonians) at low energy scales there could be emergent gauge fields and interesting consequences of quantum number fractionization, quantum order etc. 
At high energy scales gauge fields interact and it is no more simple or useful to talk in terms of emergent gauge fields. As we will see soon, the Hamiltonian invented by Kitaev on a honeycomb lattice is very special. It offers static Z$_2$ gauge fields and makes the Z$_2$ gauge theory meaningful at all energy scales. 

\subsection{Majorana fermions and the Z$_2$ theory}

We can make connection to Kitaev's representation of the spins by writing,
\begin{equation}
\label{kitrep}
c_\uparrow=\frac{c_y-ic_x}{2},~~~c_\downarrow=-\frac{c-ic_z}{2}.
\end{equation}
where $c,c_x,c_y~{\rm and}~c_z$ are Majorana fermions. The single occupancy
constraint, Eq.~(\ref{u1cons}) reduces to exactly Kitaev's form,
\begin{equation}
\label{kitcons}
cc_xc_yc_z=1.
\end{equation}
Kitaev's representation of the spins then get written as,
\begin{equation}
\label{kitsprep}
\frac{i}{2}cc_a=S^a-\tilde S^a.
\end{equation}
Note that these three operators are not equal to the gauge invariant spin 
operators in the extended Hilbert space but are exactly equivalent to them
in the physical Hilbert space. Substituting the expressions in Eq.~(\ref{kitsprep}) for the spin operators in the Hamiltonian is then equivalent
to adding gauge fixing terms. Since the $SU(2)$ gauge generators are invariant
under the Z$_2$ center of the gauge group, these terms only fix the gauge
upto the central Z$_2$ group represented by $e^{i2\pi\tilde S^3}$. The 
Hamiltonian will therefore continue to have a Z$_2$ gauge symmetry. 

The simple example of a  spin-$\frac{1}{2}$ in a magnetic field illustrates
these issues. If we take the Hamiltonian to be
\begin{equation}
\label{spbham1}
H=B~S^3.
\end{equation}
then the theory has $SU(2)$ gauge symmetry, the degenerate ground states in
the extended space are
\begin{equation}
\label{su2gs}
\vert GS\rangle=\alpha c^\dagger_\downarrow\vert 0\rangle
+\beta\vert 0\rangle
+\gamma c^\dagger_\uparrow c^\dagger_\downarrow\vert 0\rangle,
\end{equation}
for arbitrary $\alpha, \beta~{\rm and}~\gamma,
~\vert\alpha\vert^2+\vert\beta\vert^2+\vert\gamma\vert^2=1$. The last two
states in the RHS of Eq.~(\ref{su2gs}) transform as a doublet under the
$SU(2)$ gauge symmetry. Gauge averaging therefore projects out the ground 
state in the physical subspace, namely $c^\dagger_\downarrow\vert 0\rangle$. 

If the Hamiltonian is taken to be,
\begin{equation}
\label{spbham2}
H=iB~cc_z=B~\left(S^3-\tilde S^3\right).
\end{equation}
The theory has only Z$_2$ gauge invariance, the degenerate ground state in the
extended Hilbert space are,
\begin{equation}
\label{z2gs}
\vert GS\rangle=\alpha c^\dagger_\downarrow\vert 0\rangle
+\gamma c^\dagger_\uparrow c^\dagger_\downarrow\vert 0\rangle.
\end{equation}
The second state in the above equation transforms non-trivially under the
Z$_2$ gauge transformation. Thus again, under gauge averaging, the ground 
state in the physical sector is projected out.
Thus  Kitaev's representation of the spin operators can be interpreted
as adding gauge fixing terms to the $SU(2)$ gauge invariant hamiltonian 
which leave  an unbroken (unfixed) Z$_2$ gauge symmetry.
\section{The Z$_2$ gauge theory of the Kitaev model}
\label{z2gtkm}

\subsection{The Hamiltonian}

Following Kitaev \cite{kit1}, we write the hamiltonian in terms of the
Majorana fermions,
\begin{equation}
\label{z2ham1}
\tilde H=\sum_{a=1}^3\sum_{\langle ij\rangle_a}ic_iu_{\langle ij\rangle_a}c_j,
\end{equation}
where the link variables are defined as
$u_{\langle ij\rangle_a}=ic_{ai}c_{aj}$. It
is natural to express them in terms of the bond fermions \cite{short}
defined as,
\begin{equation}
\label{bfdef}
\chi_{\langle ij\rangle_a}\equiv \frac{c_{ai}+ic_{aj}}{2}.
\end{equation}
The link variables are then given in terms of the occupancy number of the bond
fermions,
\begin{equation}
\label{bfgf}
u_{\langle ij\rangle_a}=
2~\chi^\dagger_{\langle ij \rangle_a}\chi_{\langle ij\rangle_a}-1.
\end{equation}
It is easy to see that,
\begin{eqnarray}
\label{uijprop1}
u_{\langle ij\rangle_a}^2&=&1, \nonumber\\
\label{uijprop2}
[u_{\langle ij\rangle_a},H]&=&0.
\end{eqnarray}
Thus the link variables can be interpreted as static Z$_2$ gauge fields.

\begin{figure}[h!]
\center{\hbox{\epsfig{figure=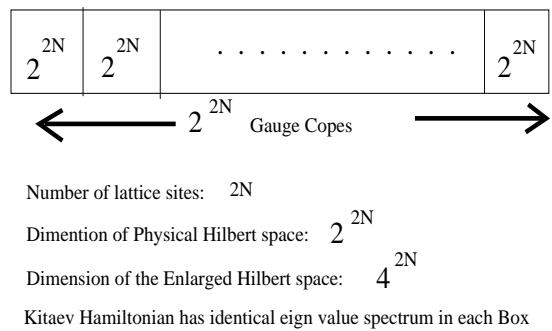,height=1.7in,width=2.9in}}}
\caption{How the enlarged Hilbert space for Kitaev Model become just Gauge Copies.}
\label{hlpt1}
\end{figure}

It is remarkable that at one shot Kitaev hamiltonian has been solved exactly for the entire many body spectrum ! Infact, two related phenomena occur:
i) the Z$_{2}$ gauge theory is exact at all energy scales and ii) the enlarged Hilbert space gets decomposed into sectors that are identical gauge copies having the same energy eigen values (Fig. \ref{hlpt1} ). The Hilbert space enlargement does not produce any unphysical state, but only gauge copies. In the standard U(1) RVB gauge theory for a Heisenberg antiferromagnet, for example, it is easy to see how unphysical states are brought in by Hilbert space enlargement. For example, an unphysical state containing M doubly occupied and M empty sites gives spectrum of a Heisenberg antiferromagnet containing 2M missing sites.

\subsection{The Jordan-Wigner transformation and Z$_2$ gauge fixing}

A remarkable feature of the Kitaev model is that the Jordan-Wigner 
transformation yields a local fermionic Hamiltonian \cite{jw1,jw2,jw3,jw4}. In this
section we show that the Jordan-Wigner transformation in the Kitaev model
can be interpreted as a Z$_2$ gauge fixing procedure resulting in gauge
invariant (gauge fixed) Majorana and bond fermions. The choice of the
Jordan Wigner path amounts to a choice of the Z$_2$ gauge.

\subsubsection{The Jordan-Wigner fermionisation}

We define the Jordan-Wigner transformation as follows ~\cite{abhinav}. Take any Hamilton
path on the lattice defined by a sequence of sites $i_n,~n=1,\dots,N_S$, 
where $N_S$ is the number of sites in the lattice. The path will classify
each bond as normal or tangential as defined in section \ref{spconsq}.
The normal bonds will form a dimer covering of the lattice.  For a given site
`$i$' we attach three vectors, two tangential vectors denoted by $\rm \hat{\bf{t}}_{1i},~ \hat{\bf{t}}_{2i}$ and one normal vectors $\rm \hat{\bf{n}}_{i}$, such that they follow Eq.~(\ref{nidef}). Then we can define the Jordon-Wigner transformations in the following compact way,

\subsubsection{Gauge invariant Jordan-Wigner fermions}

To define the Jordan-Wigner transformations we associate two Majorana fermions ($\eta_{i_n}$ and $\xi_{i_n}$) at a given site `$i_n$'. These Jordan-Wigner (Majorana) fermions are defined in terms of the gauge invariant spin-operators in the following way ,
\begin{eqnarray}
\label{etadef}
\eta_{i_n}&=&{\bf{\hat t}}_{1i_n}\cdot{\mathbf\sigma}_{i_n}~\mu_{i_n}. \\
\label{xidef}
\xi_{i_n}&=&{\bf{\hat t}}_{2i_n}\cdot{\mathbf\sigma}_{i_n}~\mu_{i_n}. \\
\mu_{i_n}&=&
\prod_{m=1}^{n-1}\left({\bf{\hat n}}_{i_m}\cdot{\mathbf\sigma}_{i_m}\right).
\label{mudef}
\end{eqnarray}
It can be easily checked that the above definitions refer to the usual anti-commutations relations for Majorana fermions,
\begin{eqnarray}
\{\eta_i,\xi_j\}&=&2\delta_{ij}, \nonumber \\
\{\xi_i,\eta_j\}&=&2\delta_{ij}, \nonumber \\
\{\eta_i,\xi_j\}&=&0 \label{jwacr1}.
\end{eqnarray}
Since the Jordan-Wigner fermions are constructed entirely from the spin 
operators, they are  manifestly gauge invariant. However, it is also 
interesting to see this by rewriting equations (\ref{xidef}) and 
(\ref{etadef}) in terms of the original Majorana fermions and gauge fields,
\begin{eqnarray}
\label{etadef1}
\eta_{i_n}&=&ic_{i_n}~\left(u_{i_{n-1}i_{n-2}}
\dots u_{i_2i_1}\right)~{\bf{\hat t}}_{1i_1}\cdot{\bf c}_{i_1}.\\
\label{xidef1}
\xi_{i_n}&=&i{\bf c}_{i_n}\cdot{\bf{\hat n}}_{i_n}
~\left(u_{i_{n-1}i_{n-2}}
\dots u_{i_2i_1}\right)~{\bf{\hat t}}_{1i_1}\cdot{\bf c}_{i_1}.
\end{eqnarray}

The transformation can be inverted to write the spins in terms of the fermions,
\begin{eqnarray}
{\bf{\hat n}}_{i_n}\cdot{\mathbf\sigma}_{i_n}&=&i\eta_{i_n}\xi_{i_n}.\\
{\bf{\hat t}}_{1i_n}\cdot{\mathbf\sigma}_{i_n}&=&\eta_{i_n}\mu_{i_n}.\\
{\bf{\hat t}}_{2i_n}\cdot{\mathbf\sigma}_{i_n}&=&\xi_{i_n}\mu_{i_n}.
\end{eqnarray}
These completes the definitions of Jordan-Wigner  fermionisations used in this article.
\subsubsection{The gauge fixed Hamiltonian}

\begin{figure}[h!]
\center{\hbox{\epsfig{figure=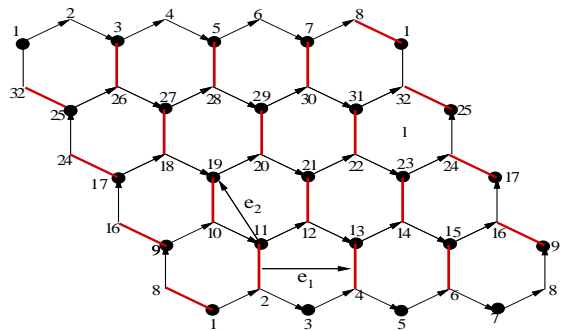,height=1.7in,width=2.9in}}}
\caption{Jordan-Wigner path on a torus for $4\times4$ lattice. The numerics at the lattice sites describe how the Jordan-Wigner path traverses the lattice.}
\label{jwpt1}
\end{figure}

The Hamiltonian can be written in terms of the gauge invariant
fermions as,
\begin{equation}
\label{gfham}
\tilde H=J_x\sum_{\langle ij\rangle_x}i\eta_i\tilde u_{ij}\eta_j
+J_y\sum_{\langle ij\rangle_y}i\eta_i\tilde u_{ij}\eta_j
+J_z\sum_{\langle ij\rangle_z}i\eta_i\tilde u_{ij}\eta_j,
\end{equation}
where the gauge fixed Z$_2$ fields, $\tilde u_{ij}$ are,
\begin{eqnarray}
\nonumber
\tilde u_{ij}&=&i\xi_i\xi_j~~~~{\rm normal~bonds}, \nonumber\\
\nonumber
&=&1~~~~~~~~{\rm tangential~bonds~except}~(ij)=(i_1i_{N_S}), \nonumber\\
\tilde u_{i_1i_{N_S}}&=&{\cal S} \nonumber\\
&=&\prod_{n=1}^{N_S}~\left({\bf{\hat n}}_{i_m}\cdot{\mathbf\sigma}_{i_m}\right) \nonumber \\
&=&
~\left(u_{i_{1}i_{N_S}} u_{i_{N_S}i_{N_S-1}}
\dots u_{i_3 i_2}u_{i_2i_1}\right) \prod_{n=1}^{N_S} \eta_{i_n}.
\label{utildedef}
\end{eqnarray}
${\cal S}$ is a well known conserved quantity in the one-dimensional
applications where it corresponds to the total number of fermions modulo 2 
and determines the boundary conditions on the fermions.
Thus the Jordan-Wigner transformation is equivalent to a gauge fixing
procedure where all the gauge fields on the tangential bonds (except one)
are set equal to 1. The choice of the Hamilton path amounts to a gauge 
choice since it defines which of the bonds are tangential. It also 
defines the sign in the definition of $u_{ij}$ for the normal bonds. In 
Eq.~(\ref{utildedef}), the sign corresponds to a Hamilton path which
winds regularly in the ${\bf{\hat e}}_1$ direction as shown in 
Fig (\ref{jwpt1}). Hereinafter all explicit computations will be with respect to 
this path. A general algorithm to go to this gauge, which we will refer to 
as the Jordan-Wigner gauge, is given in appendix \ref{gfalg}.
\subsubsection{The fermionic conserved quantities}
\label{frconsq}
All the gauge fixed Z$_2$ fields are conserved quantities. It is convenient
to express them in terms of the gauge fixed bond fermions on the normal bonds,
\begin{equation}
\chi_{ij}\equiv\frac{\xi_i+i\xi_j}{2},~~~
\chi^\dagger_{ij}\equiv\frac{\xi_i-i\xi_j}{2}.
\label{chidef}
\end{equation}
The conserved quantities are then the occupation numbers of the bond fermions,
\begin{equation}
\tilde u_{ij}=2~\chi^\dagger_{ij}\chi_{ij}-1.
\end{equation}
There are $N_p$ normal bonds and hence the bond fermion occupation numbers
form a set of $N_p$ conserved quantities. Thus along  with ${\cal S}$,
we have $N_p+1$ conserved quantities consistent with the analysis in terms
of spin variables in section \ref{spconsq}. To see the meaning of ${\cal S}$,
it is convenient to define complex fermions in the matter sector from the
two $\xi$ fermions on every normal bond,
\begin{equation}
\label{psidef}
\psi_{ij}\equiv\frac{\eta_i+i\eta_j}{2},
~~~\psi^\dagger_{ij}\equiv\frac{\eta_i-i\eta_j}{2}.
\end{equation}
It can then be shown that,
\begin{eqnarray}
{\cal S}&=&(-1)^{\left( N_\psi+N_\chi+1\right)}, \nonumber\\
N_\psi&\equiv&\sum_{\rm z~bonds}\psi^\dagger_{ij}\psi_{ij}, \nonumber\\
N_\chi&\equiv&\sum_{\rm normal~bonds}\chi^\dagger_{ij}\chi_{ij}.
\end{eqnarray}
\subsubsection{From Kitaev gauge to Jordan-Wigner gauge}
\label{kjg}
We have  explained that Jordan-Wigner gauge is a special realisation of Kitaev gauge where all the gauge fields residing on the tangential bonds are fixed to unity. One may wonder whether there exists a gauge transformation on the lattice which renders Eq.~(\ref{z2ham1}) to that of Eq.~(\ref{gfham}). Indeed there exists such a gauge transformations.  Referring to the Jordan-Wigner path given in Fig. (\ref{jwpt1}) we do the following gauge transformations at a site `$n$'

\begin{eqnarray}
c_{n} \rightarrow  \prod_{i=1,n-1} u_{i,i+1}c_{n}, ~ 1 <n <N. 
\end{eqnarray}

In the above equation the  various indices corresponds to the Jordan-Wigner path in the Fig. (\ref{jwpt1}). $u_{i,i+1}$s are the Z$_2$ gauge fields that would normally exists on the link joining site `$i$' and `$i+1$' if we apply the Majorana fermionisation as adopted by Kitaev. After implementing  the above gauge tranformations 
Z$_2$ gauge fields appear only on the normal bonds. The above gauge transformations yield  a new
conserve quantity $\mathcal{S}^{\prime}$ which appear on the bond where the Jordan-Wigner end path meets.
The expression for $\mathcal{S}^{\prime}$ is,
\begin{eqnarray}
&&\mathcal{S}^{\prime}= \prod_{i=1,N-1} u_{i,i+1}.
\end{eqnarray}
The significance of $\mathcal{S}^{\prime}$ is very similar to $\mathcal{S}$ introduced in Eq.~(\ref{utildedef}).

\section{Four fold Degeneracy on a torus}
\label{gsd}
Since the gauge fields are static, the problem is reduced to one of
non-interacting fermions on a lattice. It is known \cite{kit1, lieb}
that the lowest fermionic ground state energy is obtained for the 
flux free configuration, i.e. $B_p=1,~\forall p$. On the torus, there
are four gauge in equivalent configurations for every configurations of $B_p$ as argued in \ref{spconsq}. These
correspond to the four values of the gauge invariant conserved quantities
$W_1~{\rm and}~W_2$. First We  examine  these four different gauge field configurations for flux free sector and compute the corresponding fermionic ground state. Following this we show that this leads to four fold degeneracy of ground states in thermodynamic limit. Next we   demonstrate explicitly how to obtain the four fold degeneracy for every configurations of fluxes. To this end we derive the required operators which enables us to obtain any one of the inequivalent gauge field configurations from the other for arbitrary flux configurations.

\subsection{Degenerate ground states on a Torus}
\label{gsd1}
~~~~To start with we briefly  recapitulate the notions of Jordan-Wigner transformation and  refer to Fig. (\ref{jwpt1}). The normal bonds are the ones that form the basis of the triangular lattice except for the $(0,n,\alpha)$ line. On this line the normal bonds are the ones 
between $(0,n,a)~{\rm and}~(0,n+1,b))$. We choose the first site of the 
path to be $i_1=(0,0,b)$. The four flux free configurations are then explained
as follows. To this end we write the exact Hamiltonian and $B_p$ for this particular realisation of Jordan-Wigner transformation. We divide the Hamiltonian in three parts, $H_{int}$ , $H_{bound}$ and $H_{end}$.  $H_{int}$  includes all the internal bonds and $H_{bound}$ includes all the boundary bonds except one where the Jordan-Wigner end points meets. $H_{end}$ includes the interaction for the bond where the Jordan-Wigner end points meet each other. Similarly all the $B_{p}$'s are categorized in the above three different way. Below we write the various parts of the Hamiltonian and $B_p$'s.

\begin{eqnarray}
H_{int} &=& \sum_{m,n} iJ_{x} \eta^{a}_{m,n} \eta^{b}_{m+1,n+1} + \sum_{m,n} iJ_{y} \eta^{a}_{m,n} \eta^{b}_{m,n+1} \nonumber \\
&&
+ \sum_{m,n} iJ_{z} \tilde{u}_{m,n}\eta^{a}_{m,n} \eta^{b}_{m,n} \label{hint},
\end{eqnarray}

where $\tilde{u}_{m,n}$ is defined on each internal z-bonds.

\begin{eqnarray}
H_{bound} &=& \sum_{m,n} iJ_{y} \tilde{u}_{\substack{m,n\\m,n+1}} i\eta^{a}_{m,n}\eta^{b}_{m,n+1} \nonumber \\
&& +\sum_{m,n} J_{z} i \eta^{a}_{m,n} \eta^{b}_{m,n}. \label{hbound}
\end{eqnarray}
Where $\tilde{u}_{\substack {m,n\\ m,n+1}}$ is defined on each boundary y-bond.
%and it is given by, $(2\chi^{\dagger}_{\substack{m,n\\m,n+1}} \chi_{\substack{m%,n\\m,n+1}}-1)$.
The Hamiltonian for the end bond  is given by,
\begin{eqnarray}
H_{end}&=& -{\cal S} i\eta^{a}_{M,N} \eta^{b}_{M,N}. \label{hend}
\end{eqnarray}

%$S$ is given by,
%\begin{equation}
%{\cal S}= -\Pi_{all m,n} (i\eta^{a}_{m,n} \eta^{b}_{m,n}) U_T
%\end{equation}
Now with the definition of $\psi$ fermion and $\chi$ fermion  we get the equations (\ref{hint}), (\ref{hbound}) to be rewritten  as,

\begin{eqnarray}
H_{int} &=& \sum_{m,n} J_{x}(\psi^{\dagger}_{m,n} +\psi_{m,n})(\psi^{\dagger}_{m+1,n+1} -\psi_{m+1,n+1}) \nonumber \\
& & +J_{y}(\psi^{\dagger}_{m,n} +\psi_{m,n})(\psi^{\dagger}_{m,n+1} -\psi_{m,n+1})\nonumber\\
& &+ J_{z}\tilde{u}_{m,n}(2\psi^{\dagger}_{m,n}\psi_{m,n}-1). 
\end{eqnarray}
The Hamiltonian for the slanting bonds,
\begin{eqnarray}
H_{bound}&=& \sum_{m,n} J_{y}\tilde{u}_{\substack{m,n\\m,n+1}}(\psi^{\dagger}_{m,n} +\psi_{m,n})(\psi^{\dagger}_{m,n+1} -\psi_{m,n+1}) \nonumber \\
& &+ J_{z}(2\psi^{\dagger}_{m,n}\psi_{m,n}-1).
\end{eqnarray}
Lastly the Hamiltonian term for the end bond where the end points of the Jordan-Wigner path  meet each other is given by,
\begin{equation}
H_{end}=-{ \cal S} J_{z} (2\psi^{\dagger}_{M,N}\psi_{M,N}-1).
\end{equation}
Following equations (\ref{utildedef}), (\ref{xidef}) and (\ref{psidef}), we rewrite the complete expressions for various conserved quantities appeared in the final form of the Hamiltonian. The conserve quantity $\tilde{u}_{m,n}$ defined on each  internal z-link is given by,
\begin{equation}
\tilde{u}_{m,n}=(2\chi^{\dagger}_{m,n} \chi_{m,n}-1).
\end{equation}
Similarly the conserved quantity defined on each boundary y-bonds (which are labeled by the z-bonds it is connected with (i,e $\tilde{u}_{\substack{m,n\\m,n+1}}$))
is given by,
\begin{equation}
\tilde{u}_{\substack{m,n\\m,n+1}}= (2\chi^{\dagger}_{\substack{m,n\\m,n+1}}\chi_{\substack{m,n\\m,n+1}}-1 ).
\end{equation}
And the ${\cal S}$,  for the Jordan-Wigner gauge, is given by,
\begin{equation}
{\cal S}= -(-1)^{MN+N_{\psi}+N_{\chi}}.
\end{equation}
 If  $ \mathcal{P}^{\mathcal G}$ and $ \mathcal{P}^{\mathcal M}$ denote  the parity operators for the gauge  fermions and the matter fermions respectively, then we can write $\mathcal{S}= -(-1)^{MN}  \mathcal{P}^{\mathcal M}\mathcal{P}^{\mathcal G}$. From the Fig. \ref{jwpt1}, we see that a single hexagon always contains two normal bonds where  each normal bond is associated with a conserved(static) Z$_2$ gauge field . $B_p$ for any plaquette is  the product of these two conserved Z$_2$ gauge fields. Thus, $B_p = \tilde{u}_{ij} \tilde{u}_{kl}$ ,  where `$ij$' and `$kl$' are the normal bonds for the  plaquette `$p$'. For the end plaquette where Jordan-Wigner path terminates $B_p$ is given by $B_p = - {\cal S}\tilde{u}_{ij} \tilde{u}_{kl}$. \\
Now we are in a position to show the  ground state degeneracy in thermodynamic limit. To this end we explicitly write the four in-equivalent gauge field configurations corresponding the flux free configurations and write down the corresponding fermionic Hamiltonian. Finally we find the spectra for each of these four fermionic Hamiltonian. 
\subsubsection {Choice 1}
Here the flux free configuration is obtained by making all  the $\tilde{u}$'s to be 1 and ${\cal S}=-1$. The loop  conserve quantities are having the following eigenvalue ,$W_{1}=1$ and $W_{2}=1 $. These particular choice makes the resulting fermionic Hamiltonian  transitionally invariant and usual Periodic boundary condition in both  the direction can be used to diagonalize the Hamiltonian. We will explicitly write the Hamiltonian in terms of complex fermion $\psi$. However one can equivalently work in terms of $\eta$, Majorana fermion representation. To keep this in mind we will continue to mention appropriate gauge transformations for $\eta$ fermions as well as $\psi$ fermions.
%\begin{eqnarray}
%H&=& \sum_{m,n} iJ_{x} \eta^{a}_{m,n} \eta^{b}_{m+1,n+1} + \sum_{m,n} iJ_{y} \eta^{a}_{m,n} \eta^{b}_{m,n+1} \nonumber \\
%& &
%+ \sum_{m,n} iJ_{z}\eta^{a}_{m,n} \eta^{b}_{m,n} 
%\end{eqnarray}
The complete translational invariant Hamiltonian is given by,
\begin{eqnarray}
H &=& \sum_{m,n} J_{x}(\psi^{\dagger}_{m,n} +\psi_{m,n})(\psi^{\dagger}_{m+1,n+1} -\psi_{m+1,n+1}) \nonumber \\
& & +J_{y}((\psi^{\dagger}_{m,n} +\psi_{m,n}))(\psi^{\dagger}_{m,n+1} -\psi_{m,n+1})\nonumber\\
& &+J_{z}(2\psi^{\dagger}_{m,n}\psi_{m,n}-1).
\end{eqnarray}

This is a manifestly p-wave superconducting Hamiltonian and can  be easily diagonalized by going to the momentum space.
%The boundary condition on $\eta_{m,n}$ is $\eta_{M+1,n}=\eta_{1,n}$ and $\eta_{m,N+1}=\eta_{m,1} $ , or alternatively $\psi_{M+1,n}=\psi_{1,n}$ and $\psi_{m,N+1}=\psi_{m,1}$. However we prefer to work in the representation of $\psi$ where
%implementing the boundary condition on fermion occupation number is easy. 
The constraint on the number of $\psi$ fermion becomes,
\begin{equation}
\prod (2\psi^{\dagger}_{m,n}\psi_{m,n}-1)=1,
\end{equation}
which implies that  we are to fill up only the even number of $\psi$ fermions. We  define the Fourier transform of the $\psi_{m,n}$ as given below,
\begin{equation}
\psi_{m,n}=\frac{1}{\sqrt{MN}}\sum_{p,q} e^{i\left(k_{1}m+k_{2}n\right)} \label{ftc1},
\end{equation}
where $k_{1}= 2\pi\frac{p}{M} $ and $k_{2}= 2\pi\frac{q}{N}$. 
This is obtained by noticing the fact that we can write ${\vec{k}}=\frac{p}{M}{\vec{G_1}}+\frac{q}{N}{\vec G_2}$ where ${\vec G_{1/2}} $ are the reciprocal lattice vectors which are  given by,
\begin{equation}
{\vec{G_{1}}}= \frac{4\pi}{\sqrt{3}}\left(\frac{\sqrt{3}}{2} \textbf{e}_{x} +\frac{1}{2} \textbf{e}_{y}\right) \,\,;\,\, {\vec{G_{2}}} = \frac{4\pi}{\sqrt{3}}  \textbf{e}_{y}. 
\end{equation}
Substituting this  we get the resulting Hamiltonian in momentum space as,
\begin{eqnarray}
H &=&\sum_{k} (\epsilon_{k} \psi^{\dagger}_{k} \psi_{k} - \epsilon_{k} \psi_{-k} \psi^{\dagger}_{-k} +i \delta_{k} \psi^{\dagger}_{k} \psi^{\dagger}_{-k} -i \delta_{k} \psi_{-k} \psi_{k})\nonumber\\
&&+ \epsilon_{0,0} \psi^{\dagger}_{0,0}\psi_{0,0} + \epsilon_{\pi,0} \psi^{\dagger}_{\pi,0}\psi_{\pi,0}+\epsilon_{0,\pi} \psi^{\dagger}_{0,\pi}\psi_{0,\pi}\nonumber\\
& &+\sum_{k} \epsilon_{k} -MN J_z \label{1ckm}.
\end{eqnarray}
%\begin{eqnarray}
%H=\sum(\epsilon_{k}\psi^{\dagger}_{k}\psi_{k} +i\frac{\delta_{k}}{2}\psi^{\dagger}_{k}\psi^{\dagger}_{-k} -i\frac{\delta_{k}}{2}\psi_{-k}\psi_{k} )
%\end{eqnarray}

 Where $\epsilon_{k}=2(J_x \cos k_x +J_y \cos k_y +J_z) $ and 
 $ \delta_k=2(J_x \sin k_x +J_y \sin k_y )$. 
 %With $k_{x}=k_{1}+k_{2} ,k_{y}=k_{2}$. 
 %$k_1 \rm and k_2$ are the reciprocal  vector in the direction of $\bf{e_1} ,\,\,\rm \bf{e_2}$ respectively.
%$k_{1}= 2\pi\frac{m}{M} $ and $k_{2}= 2\pi\frac{n}{N}$. 
$k_{x}=\textbf{k}.\textbf{n}_x$, $k_{y}=\textbf{k}.\textbf{n}_y$ and 
$\textbf{n}_{x,y}=\frac{1}{2}\textbf{e}_{x} \pm
\frac{\sqrt{3}}{2}{\textbf{e}_{y}}$ are unit vectors along x and y type bonds.
%Now we rewrite the Hamiltonian in the following way,
%\begin{eqnarray}
%H &=&\sum_{k} (\epsilon_{k} \psi^{\dagger}_{k} \psi_{k} - \epsilon_{k} \psi_{-k} \psi^{\dagger}_{-k} +i \delta_{k} \psi^{\dagger}_{k} \psi^{\dagger}_{-k} -i \delta_{k} \psi_{-k} \psi_{k})\nonumber\\
%&&+ \epsilon_{0,0} \psi^{\dagger}_{0,0}\psi_{0,0} + \epsilon_{\pi,0} \psi^{\dagger}_{\pi,0}\psi_{\pi,0}+\epsilon_{0,\pi} \psi^{\dagger}_{0,\pi}\psi_{0,\pi}\nonumber\\
%& &+\sum_{k} \epsilon_{k} -MN J_z 
%\end{eqnarray}

In Eq.~(\ref{1ckm}), the sum over `$k$' runs over first half of the Brillouin zone and does not
include the `$k$'-points $(\pi,0), (0,\pi),(0,0)$. The first line of the Hamiltonian
is diagonalize by the following transformations,

\begin{equation}
\left( \begin{array}{r}
\alpha_{k}\\
\beta_{k} \end{array}\right)=\left( \begin{array}{rr} 
\rm cos \theta_{k} & -\rm i \;\rm sin\theta_{k} \\
\rm -i \; sin \theta_{k} & \rm cos \theta_{k} \\
\end{array} \right) \left( \begin{array}{r}
\psi_{k}\\
\psi^{\dagger}_{-k} \end{array}\right).
\end{equation}
Where $\cos2\theta_{k}=\epsilon_{k}/E_{k}$, with $E_{k}=\sqrt{\epsilon^2_k + \delta^2_k}$.
Then re-witting the Hamiltonian we get,
\begin{eqnarray}
H&=&\sum E_{k}(\alpha^{\dagger}_{k}\alpha_{k}-\beta^{\dagger}_{k}\beta_{k})\nonumber\\
&&  +\epsilon_{0,0} \psi^{\dagger}_{0,0}\psi_{0,0} + \epsilon_{\pi,0} \psi^{\dagger}_{\pi,0}\psi_{\pi,0}+\epsilon_{0,\pi} \psi^{\dagger}_{0,\pi}\psi_{0,\pi} \nonumber\\
&&-\frac{1}{2}(\epsilon_{0,0}+\epsilon_{0,\pi}+\epsilon_{\pi,0}) +(\sum_{k^{\prime}}\frac{1}{2}\epsilon_{k^{\prime}}-N_z J_z).
\end{eqnarray}
Here sum over $k^{\prime}$ runs over full Brillouin zone. The last term in the parenthesis is always zero for Torus. 

\subsubsection{Choice 2}
 Here the flux free configuration is obtained by making all $\tilde{u}$'s  -1 and ${\cal S}=-1$. The corresponding values of loop conserve quantity is given by,$W_1=-1$ and $W_2=-1$. To implement the Fourier transformation we do the following steps.
We make the following gauge transformation. $\eta^{b}_{M,n} = -\eta^{b} _{M,n}$ for all `$n$'.  In terms of $\psi$ fermion the necessary gauge transformation is $\psi_{M,n}=-\psi^{\dagger}_{M,n}$ for all `$n$'.  Then the Hamiltonian requires  $\eta_{M+1,n}=-\eta_{1,n}$ and $\eta_{m,N+1}=\eta_{m,1}$(or alternatively $\psi_{M+1,n}=-\psi_{1,n}$ and $\psi_{m,N+1}=\psi_{m,1}$).  This is equivalent to   anti-periodic boundary Condition in $\bf e_{1}$ direction and periodic boundary condition in  $\bf e_{2}$ direction.  The necessary Fourier transform is defined with,
\begin{equation}
\psi_{m,n}=\frac{1}{\sqrt{MN}}\sum_{p,q} e^{i\left(k_{1}m+k_{2}n\right)},
\end{equation}
with $k_{1}= \frac{2\pi}{M}(m+\frac{1}{2}); \,\,\,\, k_{2}= 2\pi\frac{n}{N}$.
Substituting this in the Hamiltonian and diagonalizing straightforwardly we get,
\begin{eqnarray}
H&=&\sum_k E^{2}_{k}(\alpha^{\dagger}_{k}\alpha_{k}-\beta^{\dagger}_{k}\beta_{k}) +(\sum_{k^{\prime}}\frac{1}{2}\epsilon_{k^{\prime}}-N_z J_z) \label{2ckm}.
\end{eqnarray}

Here also `$k$' runs over first half of the Brillouin zone and the `$k^{\prime}$' runs over the full Brillouin zone. Note the absence of $(0,\pi),(\pi,0) \rm \, (0,0)$ mode. They do not appear here for this anti-periodic boundary condition. This will be true for the choices 3 and 4 also. Various parameters appearing in Eq.~(\ref{2ckm}) are given below,
\begin{eqnarray}
&&E^{2}_{k}=\sqrt{(\epsilon_{k}^{2}+\delta^{2}_{k})}, \nonumber \\
&& \epsilon_{k}=2(J_{x} \cos k_{x} +J_{y}\cos k_{y} -J_{z}), \nonumber \\
&& \delta_{k}=2(J_{x}\sin k_{x}+J_{y}\sin k_{y}).
\end{eqnarray}
\subsubsection{Choice 3}
For this case the flux free configuration is obtained by making  $\tilde{u}_{M,n}=-1 \rm $ where `$n$' runs from 1 to $N-1$. All other $\tilde{u}$'s are 1 and ${\cal S}=1$. The loop conserve quantity $W_1$ takes value 1 and and $W_2$ -1.  Similar to the previous case we need to do following gauge transformations in order to apply Fourier transform. We make   $\eta^{b}_{m,N} = -\eta^{b} _{m,N}$ for all `$m$'.  In terms of $\psi$ fermion the necessary gauge transformation is $ \psi_{m,N}=-\psi^{\dagger}_{m,N}$ for all `$m$. The resulting Hamiltonian requires  $\eta_{m,N+1}=-\eta_{m,1}$ and $\eta_{M+1,n}=\eta_{1,n}$( $\psi_{m,N+1}=-\psi_{m,1}$ and $\psi_{M+1,n}=\psi_{1,n}$). This indicates  anti-periodic boundary condition in $\bf e_{2}$ direction and periodic boundary condition in  $\bf e_{1}$ direction. The resulting Fourier transform is defined with,
\begin{equation}
k_{1}= 2\pi\frac{m}{M}; \,\,\,\,\, k_{2}= \frac{2\pi}{N}(n+\frac{1}{2}).
\end{equation}
The resulting Hamiltonian in `k' space is similar to Eq.~(\ref{2ckm}) with  $\epsilon_{k}=2(J_{x} \cos k_{x} +J_{y}\cos k_{y} +J_{z})$. 

\subsubsection{Choice 4}
For this choices we need, $\tilde{u}_{M,n}=1 \rm$ where `$n$' runs from 1 to $N-1$ . All other $\tilde{u}$'s are -1 and ${\cal S}=1$. $W_1=-1$ and $W_2=1$. In this case  one requires combined gauge transformation mentioned for the choices 2 and 3. This makes the Hamiltonian  anti periodic  in both the directions. The required Fourier transform is defined with,
\begin{equation}
k_{1}= \frac{2\pi}{M}(m+\frac{1}{2}) ;\,\,\,\, k_{2}= \frac{2\pi}{N}(n+\frac{1}{2}).
\end{equation}
Proceeding as before we get exactly Eq.~(\ref{2ckm}) with identical expression for $\epsilon_{k}$.

\subsubsection{Ground State Energy in  thermodynamic limit}
To get the ground state energy for the gauge choices 2,3 and 4, one  fills up the negative energy states consistent with the boundary condition (i.e, to satisfy the
constraint ${\cal S}$ which restricts the total number of particles (odd or even number) to be taken). In the limit of $M ,N \rightarrow \infty $ all the above four choices, the ground state energy is obtained as,
\begin{equation}
E_{G}= \frac{\sqrt{3}}{16\pi^{2}}\int_{BZ} E(k_{x},k_{y})dk_{1}dk_{2},
\end{equation}
with $E(k_{x},k_{y})$ is defined before. The appearance of a `-' sign in the expression of $\epsilon_{k}$ for  choices 2 and 4 can be accounted for by shifting the $k_1$ integral to $\pi - k_1$. Thus it is clear that in thermodynamic limit ground state has four fold  degeneracy. 
\subsection{Four fold Topological degeneracy for any eigenstate}
\label{tpg}

In the preceding section, we have shown that four degenerate ground states  are
characterized by four different topologically distinct gauge field
configurations. From the section \ref{z2gtkm}, we  infer that every flux
configuration is  characterized by such four topologically distinct gauge field
configurations. This leads to a four-fold degeneracy for every eigenstate,
including the  ground state, in the thermodynamic limit. 

We now demonstrate this explicitly. Our proof is similar to that of 
Wen and Niu \cite{wen-niu} which shows the topological degeneracy of
fractional quantum Hall states on a torus. We construct two operators which 
we call $V_1$ and $V_2$, that act on states with a given gauge field 
configuration and produce states with a different gauge field configuration 
without changing the values of the flux operators, $B_p$. They however change 
the values of the Wilson loop operators, $W_{1,2}$. These two operators are 
therefore the generators of the so called ``large gauge transformations".
There are four topologically different sectors of gauge field configurations
corresponding to $W_i=\pm 1$.
We further show that only effect of the large gauge transformations on the 
matter sector is to change the boundary conditions of the Majorana fermions
from periodic to anti-periodic or vice versa. Thus the energy eigenvalues
only change by $\sim 1/L$, where $L$ is the length of the torus. The 
eigenstates in the four sectors (related by the action of $V_i$) are therefore
degenerate in the thermodynamic limit.
 
The four operators $V_i,~W_i$ characterise the topological degeneracy.
We show that $V_i$ and $W_i$ satisfy the following algebra,
\begin{eqnarray}
\{V_1,V_2\}=0,~~ \{V_i,W_i\}=0,~~ [V_i,W_j]_{i\ne j}=0.\label{cmtr}
\end{eqnarray}
We can then construct four operators, $T_a,~a=1,\dots,4$,
\begin{eqnarray}
\label{tadef}
T_1=V_1 W_1, ~~ T_2=V_2 W_2,~~ T_3=V_1, ~~ T_4=V_2.
\end{eqnarray} 
These satisfy the Clifford algebra,
\begin{equation}
\label{cliff}
\{T_a,T_b\}=2\delta_{ab}
\end{equation}
Thus we show that the four-fold topological degeneracy on the torus
is characterised by the 4-dimensional Clifford algebra.

We will first write  down and discuss the expressions for $V_1$ and $V_2$
for the 32 site system illustrated in Fig. \ref{jwpt1}. The construction
easily generalises for any even-even lattice.
\begin{eqnarray}
\label{visdef}
V_1&=&\sigma^z_1 \sigma^z_3 \sigma^z_5 \sigma^z_7\\ 
V_2&=&\sigma^y_1\sigma^y_{16}\sigma^y_{17}\sigma^y_{32}
\prod^7_{i=2}\sigma^z_i\prod^{23}_{i=18}\sigma^z_i
\end{eqnarray}
The two Wilson loops for this lattice are,
\begin{eqnarray}
\label{w1s}
W_1&=&
\sigma^y_1\sigma^y_8\sigma^y_9\sigma^y_{16}\sigma^y_{17}
\sigma^y_{24}\sigma^y_{25}\sigma^y_{32}\\
W_2&=&\prod_{i=1}^8\sigma^z_i
\end{eqnarray}
It can be verified that the above constructions satisfy the algebra
in equation (\ref{cmtr}) and hence the topological operators defined
in equation (\ref{tadef}) satisfy the Clifford algebra (\ref{cliff}).
It can also be verified that $V_i$ commute with all the $B_p$'s.

Now consider simultaneous eigenstates of the Wilson loop operators,
\begin{equation}
\label{westates}
W_i\vert w_1,w_2\rangle=w_i\vert w_1,w_2\rangle~,~~~~~~~~w_i=\pm 1
\end{equation}
The algebra in equation (\ref{cmtr}) implies,
\begin{equation}  
\label{vonwes}
V_1\vert w_1,w_2\rangle=\vert -w_1,w_2\rangle,~~
V_2\vert w_1,w_2\rangle=\vert w_1,-w_2\rangle
\end{equation}
Thus we have shown that $V_i$ are the generators of large gauge
transformations.

We next consider their action on the hamiltonian.
\begin{equation}
\label{vonh}
V_1HV_1^{-1}=H^1,~~~~~~V_2HV_2^{-1}=H^2
\end{equation}
where $H^1$ is the same as $H$, except that the bonds on
one non-trivial loop in the ${\bf e}_1$ direction have changed sign,
namely the loop $(1,2,3,4,5,6,7,8,1)$. In $H^2$, a line of parallel
bonds in the ${\bf e}_1+{\bf e}_2$ direction have changed sign. Namely,
the bonds $(7-8),~(31-32),~(23-24)$ and $(16-17)$. We will now write down 
the operators for a general even-even lattice and then show that the 
transformed hamiltonians are degenerate in the thermodynamic limit. We
will show that in the fermionised theory, these changes of sign can be 
absorbed into the single particle eigenfunctions of the Majorana fermions
and change the single particle energy eigenvalues by $\sim 1/L$. Thus the 
energies do not change in the thermodynamic limit, making every 
many body eigenstate four-fold degenerate.

The general expressions for $V_{1(2)}$ are given by
\begin{eqnarray} 
&&V_1= \prod^{M}_{m=1} \sigma^{b,z}_{m,0},
\nonumber \\ 
&&V_2= \prod^{N/2} _{n=1} \sigma^{a,y}_{0,2n}  \sigma^{b,y}_{0,2n}
\prod^{m=1,M-1}_{n=1,N/2} \sigma^{a,z}_{m,2n-1} \sigma^{b,z}_{m,2n-2} \label{v1v2}.
\end{eqnarray}
It can be verified that these constructions satisfy the algebra in equation
(\ref{cmtr}) and also commute with all the $B_p$'s. In $H^1$, the bonds
$(m,0,b)-(m+1,1,a) ~ {\rm{and}}~ (m,1,a)-(m,0,b)$ change sign and in $H^2$, the bonds $(M-1,n,b)-(0,n+1,a)$
change sign.

In the fermionised theory, the single particle eigenfunctions satisfy the 
equation,
\begin{equation}
\label{spe1}
\sum_jiA_{ij}\phi^n_j=\epsilon^n\phi^n_i
\end{equation}
where $A_{ij}$ is an antisymmetric matrix coupling the nearest neighbours
of the honeycomb lattice. The eigenvalues come in pairs and we denote
$(\phi^n)^*=\phi^{-n}$, $\epsilon^{-n}=-\epsilon^n$. $n$ will then go 
from $1,\dots,NM$. The Hamiltonian is diagonal in terms of the complex 
fermions defined by,
\begin{eqnarray}
\label{alphadef}
\alpha_n&=&\sum_i\phi^n_{ia}\eta_i^a+\phi^n_{ib}\eta_i^b\\
\label{diagham}
H&=&\sum_n\epsilon^n\left(2\alpha^\dagger_n\alpha_n-1\right)
\end{eqnarray}
We now make the transformation,
\begin{equation}
\label{philgt1}
\phi^{n\prime}_{nm,a}= e^{in\pi/N}\phi^n_{nm,a}
\end{equation}
Equation (\ref{spe1}) then gets written as,
\begin{equation}
\label{spe2}
\sum_j A_{ij} \phi^n_j=\sum_j\left(iA^1_{ij}+i\delta A_{ij}\right)\phi^{n\prime}_i
=\epsilon^n\phi^{n\prime}_i
\end{equation}
where $A^1$ is the antisymmetric matrix corresponding to $H^1$ and
$\delta A_{ij}\propto 1/N$ when $N$ is very large. Thus the single
particle energy eigenvalues of $A$ and $A^1$ are identical in the 
thermodynamic limit when $N \rightarrow \infty$. The spectrum of $H$ and $H^1$ are also therefore
identical. The mapping of the eigenvalues of $H$ and $H^2$ can be similarly
shown using the transformation,
\begin{equation}
\label{philgt2}
\phi^{n\prime}_{nm,a}= e^{im\pi/M}\phi^n_{nm,a}
\end{equation}

We can also write equation (\ref{v1v2}) in terms of the gauge invariant 
fermions. $V_1$ or $V_2$ can be written as $V_i=V^{\mathcal M}_i V^{\mathcal G}_i$ ($i=1,2$). As the
notation suggests, $V^{\mathcal M}_i$'s and $V^{\mathcal G}_i$'s are composed of operators which 
belong to the matter sector and the gauge sector respectively. They are 
given by, 
\begin{eqnarray}
&&V^{\mathcal G}_1=\prod \xi^a_{m,0}, ~ m=1, M-1, \nonumber \\
&&V^{\mathcal M}_1=\prod \eta^{a}_{m,0}, ~m=1,M, \nonumber \\
&&V^{\mathcal G}_2= \prod^{N/2} _{n=1} \xi^{a}_{0,2n}  \xi^{b}_{0,2n} 
\prod^{j=1,N/2}_{k=2,N} \xi^{a}_{m,2n-1} \xi^{b}_{m,2n-2}, \nonumber  \\
&&V^{\mathcal M}_2= \prod^{N/2} _{n=1} \eta^{a}_{0,2n}  \eta^{b}_{0,2n} 
\prod^{m=1,M-1}_{n=2,N/2} \eta^{a}_{m,2n-1} \eta^{b}_{m,2n-2}.
\end{eqnarray}

From the above expressions  we  can easily find the commutation relations of parity operator for gauge fermions and the matter fermions. We find that,
\begin{eqnarray}
&&\{\mathcal{P}^{G}, V_1\}=0,~[\mathcal{P}^{\mathcal G}, V_2]=0,~[\mathcal{P}^{\mathcal M}, V_1]=0,~[\mathcal{P}^{\mathcal M}, V_2]=0. \nonumber \\
&&
\end{eqnarray}

The fact that the  parity of the matter fermions are conserved is consistent with the fourfold degeneracy   discussed here and in  (\ref{gsd1}).
\section{Discussion} 
\label{dis} 

In this paper we have  discussed many important aspects of Kitaev model. We
have shown how the $SU2$ gauge contained in the Kitaev model and explained
explicitly how this  $SU2$ gauge symmetry is reduced to Z$_2$ gauge symmetry.
We have solved the Kitaev model using Jordan-Wigner method in a general way.
Though Jordan-Wigner transformations has been used earlier to solve for Kitaev
model, our formalism reveals many new features, for example, our definition of
J-W transformation is applied to a torus. We  focused on the gauge field
contents of the Kitaev model and the issue of topological degeneracy. We showed
that ground state is four fold degenerate on the torus {\em in both phases}.
While it indicates non-trivial topological order, the ground state degeneracy
does not distinguish between the gapless and gapped phases. Finally we have
shown the equivalence between the fermionised Hamiltonian obtained in Kitaev
gauge and J-W gauge. Lastly we have constructed four mutually anti-commuting
operators on a torus to illustrate explicitely the four fold degeneracy for
every eigenstate.  Our analysis reveals that Jordan-Wigner analysis can  be
used even in quantum spin liquid problems, as a general method, to bring out
non trivial gauge field content, thereby providing an alternative method in
resonating valence bond theories.

\section{Acknowledgements}
As we were completing our manuscript, we learned of a very recent paper by
F. J. Burnell and Chetan Nayak (Ref. \cite{nayak}), which has some overlap with the present paper. GB thanks F. J. Burnell for
 reference[40]. This research was partly supported by Perimeter Institute for Theoretical Physics.
\appendix 
\section{The gauge fixing algorithm} 
\label{gfalg}

In this appendix we detail the gauge fixing algorithm to go from any gauge
field configuration to one in the Jordan-Wigner gauge.  First we will write
Hamiltonian as obtained from the Eq.~(\ref{gfham}). We notice that for $m\ne
0$,
\begin{eqnarray}
{\bf \hat{t}}^a_{1m,n}=\hat{y};~~ {\bf \hat{t}}^a_{2m,n}=\hat{x}; ~~{\bf \hat{n}}^a_{1m,n}=-\hat{z}, \nonumber\\
{\bf \hat{t}}^b_{1m,n}=\hat{x}; ~~{\bf \hat{t}}^b_{2m,n}=\hat{y}; ~~{\bf \hat{n}}^a_{1m,n}=\hat{z}.
\end{eqnarray}
And for  $m=0$,
\begin{eqnarray}
{\bf \hat{t}}^a_{1m,n}=\hat{z}; ~~{\bf \hat{t}}^a_{2m,n}=\hat{x};~~ {\bf \hat{n}}^a_{1m,n}=-\hat{y}, \nonumber\\
{\bf \hat{t}}^b_{1m,n}=\hat{x};~~ {\bf \hat{t}}^b_{2m,n}=\hat{z};~~ {\bf \hat{n}}^b_{1m,n}=-\hat{y}.
\end{eqnarray}
Then starting with the Ferromagnetic Hamiltonian we get, for $m=0,n=0$, 
\begin{equation}
-\sigma^{az}_{0,0} \sigma^{bz}_{0,0}= -i\eta^{a}_{0,0}\eta^{b}_{0,0}{\cal S}.
\end{equation}

For  $m\ne0$
\begin{equation}
-\sigma^{az}_{m,n} \sigma^{bz}_{m,n}= -i\eta^{a}_{m,n}\eta^{b}_{m,n}( i\xi^{a}_{m,n}\xi^{b}_{m,n} ).
\end{equation}
now for  $m=0,n\ne0$
\begin{equation}
-\sigma^{az}_{0,n} \sigma^{bz}_{0,n}= i\eta^{a}_{0,n}\eta^{b}_{0,n}.
\end{equation}

The above three equations gives complete description of all the z-z interaction.
Now for the y-bond we get for $m=0$
\begin{equation}
-\sigma^{ay}_{0,n} \sigma^{by}_{0,n+1}= -i\eta^{a}_{0,n}\eta^{b}_{0,n+1}( i\xi^{a}_{0,n}\xi^{b}_{0,n+1} ).
\end{equation}
For $m\ne0$ we get
\begin{equation}
-\sigma^{ay}_{m,n} \sigma^{by}_{m,n+1}= i \eta^{a}_{m,n} \eta^{b}_{m,n+1}.
\end{equation}
At last we write Hamiltonian for x-interaction. For $m=0$
\begin{equation}
-\sigma^{ax}_{0,n} \sigma^{bx}_{1,n+1}= -i\eta^{a}_{0,n}\eta^{b}_{1,n+1}.
\end{equation}
And for $m\ne0$ we get,
\begin{equation}
-\sigma^{ax}_{m,n} \sigma^{bx}_{m+1,n+1}= -i \eta^{a}_{m,n} \eta^{b}_{m+1,n+1}. 
\end{equation}

Now we make the following gauge transformation which we call Jordan-Wigner gauge.
\begin{equation}
\eta^{r}_{m,n}\rightarrow (-1)^{m}\eta^{r}_{m,n}.
%\,\, ;\,\,\xi^{r}_{m,n}\rightarrow (-1)^{m}\xi^{r}_{m,n} 
\end{equation}

Then we choose for each normal bond $\tilde{u}_{i,j}= -i\xi^{a}_i \xi^{b}_j$. This gives the Jordan-Wigner Hamiltonian given by equations \ref{hint}, \ref{hbound} and \ref{hend}. Then we define complex fermion $\chi_{m,n}$ on each  normal internal z-bond in the following way,
\begin{equation}
\xi^{a}_{m,n}= (\chi_{m,n}+\chi^{\dagger}_{m,n}) \,\,;\,\,\xi^{b}_{m,n}=\frac{-1}{i} (\chi_{m,n}-\chi^{\dagger}_{m,n}).
\end{equation}
Similarly on each normal slanted y-bond which is joined with a z-link $(m,n)$ and $(m,n+1)$ we define,
\begin{equation}
\xi^{a}_{m,n+1}= (\chi_{\substack{m,n\\m,n+1}}+\chi^{\dagger}_{\substack{m,n\\m,n+1}}) \,\,;\,\,\xi^{b}_{m,n}= \frac{-1}{i}(\chi_{\substack{m,n\\m,n+1}}-\chi^{\dagger}_{\substack{m,n\\m,n+1}}).
\end{equation}
With this we have alway $\tilde{u}_{i,j}=(2\chi^{\dagger}_{i,j} \chi_{i,j}-1)$. At last on each z-link we define $\psi$ in the following way,
\begin{equation}
\eta^{a}_{m,n}=(\psi_{m,n}+\psi^{\dagger}_{m,n}) \,\,;\,\,\eta^{b}_{m,n}=\frac{1}{i} (\psi_{m,n}-\psi^{\dagger}_{m,n}).
\end{equation}
Now the quantity ${\cal S}$  for lattice of dimension $(M,N)$, is given by, ${\cal S}= -(-1)^{MN+N_{\psi}+N_{\chi}}$. Here $ N_{\psi}$ and $N_{\chi}$ are the number of $\psi$ and $\chi$ fermions respectively. Now noticing the number of gauge transformations needed
for various flux  free configuration we find that we need to fill even number of $\psi$ fermions for each different gauge choices. All the results derived here are based on a representative lattice of dimensions $M$ and $N$ where $M$ and $N$ are both even. It is straight forward to carry the analogous calculation for lattice where $M$ and $N$ can be anything, odd or even. However the results obtained here should not change in thermodynamic limit.
\\

%\begin{tabular}{|l|l|l|l|l|}\hline \hline
%&&
%\multicolumn{3}{c|}{Number of $N_{\psi}$}\\\hline
%
%Flux &Expression for ${\cal S}$& M even& M even&M odd  \\
%Configuration & &N even&N odd&N odd\\\hline
%Choice1& $(-1)^{N_{\psi}+1}$&even&even&even\\ \hline
%Choice2&$(-1)^{MN+N+N_{\psi}+1}$ &even&odd&even\\ \hline
%Choice3&$(-1)^{M-N+N_{\psi}}$ &even&odd&even\\ \hline
%Choice4& $(-1)^{MN+N+N_{\psi}}$&even&odd&even\\ \hline\hline
%\end{tabular}
%\\
%Though in the thermodynamic limit one can neglect this constraints.


\begin{thebibliography}{99}
\bibitem{qcgen1} M. A. Nielsen and I. L. Chang, {\em Quantum Computation and Quantum Information} (Cambridge University Press, Cambridge, England 2000).
.\bibitem{qcgen2} C. H. Bennett and D. P. DiVincenzo, Nature {\bf 404} 247(2000).
\bibitem{qcgen3} A. Yu. Kitaev, A. H. Shen and M. N. Vyalyi, {\em Classical and Quantum Computation} (Americal Mathematical Society, 2002). 
\bibitem{qcgen4}S. Das Sarma, M. Friedman and C. Nayak, Phys. Today, {\bf 59} 32(2006).
\bibitem{qcgen5} J. Preskill, Phys. Today, {\bf 52} 24(1999).

\bibitem{kit1}
 A.Yu. Kitaev, \textit{Annals of Physics}, {\bf 321}, (2006), 2-111.

\bibitem{kit2} 
A.Yu.Kitaev, \textit{Annals of Physics}, \textbf{303}, (2003), 2-30.

\bibitem{short} 
G. Baskaran, Saptarshi Mandal, R Shankar, Phys. Rev. Lett, {\bf 98}, 247201(2007).

\bibitem{rvb-dm1} X. G. Wen, F. Wilczek, and A. Zee, Phys. Rev. B {\bf 39}, 11413(1989). 
\bibitem{rvb-dm2} N. Read and S. Sachdev, Phys. Rev. Lett. {\bf 66}, 1773(1991).\bibitem{rvb-dm3} D. S. Rokhsar and S. A. Kivelson, Phys. Rev. Lett. {\bf 61}, 2376(1988).
\bibitem{rvb-dm4} N. Read and B. Chakraborty, Phys. Rev. B {\bf 40}, 7133(1989). 
\bibitem{rvb-dm5} R. Moessner and S. L. Sondhi, Phys. Rev. Lett. {\bf 86}, 1881(2001).

\bibitem{frac1} V. Kalmayer and R. B. Laughlin, Phys. Rev. Lett. {\bf 59}, 2095(1987).
\bibitem{frac2} Senthil. T, Fisher. M.P.A, Phys. Rev. B {\bf 62}, 7850(2000).
\bibitem{jw1}Xiao-Yong Feng, Guang-Ming Zhang, Tao Xiang, Phys. Rev. Lett. {\bf 98}, 087204(2007).
\bibitem{jw2} Han-Dong Chen and Jiangping Hu, Phys. Rev. B {\bf 76}, 193101(2007).
\bibitem{jw3} Han-Dong Chen and Zohar Nussinov, \textit{J. Phys. A: Math. Theor.} {\bf 41}, 075001, 2008.
\bibitem{jw4} Saptarshi Mandal, R Shankar, G Baskaran, Poster 14, International Conference on Mott Insulator, IISc, Bangalore, July,2006,http://www.physics.iisc.ernet.in/seminars-past.php.
\bibitem{othkit1} K. P. Schmidt, S Dusuel and J. Vidal, Phys. Rev. Lett. {\bf 100}, 057208(2008).
\bibitem{othkit2} K. Sengupta, D. Sen, and S. Mondal, Phys. Rev. Lett. {\bf 100}, 077204(2008).
\bibitem{othkit3} G. Baskaran, D. Sen and R. Shankar, Phys. Rev. B {\bf 78}, 115116(2008).
\bibitem{othkit4} S. Mandal, N Surendran, Phys. Rev. B {\bf 79}, 024426(2009).
\bibitem{othkit5} S. Dusuel, K. P. Schmidt, J. Vidal, Phys. Rev. Lett. {\bf 100}, 177204(2008).
\bibitem{bza} G. Baskaran, Z. Zou and P. W. Anderson, Solid St. Commn. 63973(1987)
\bibitem{su21} Ian Affleck, Z. Zou, T. Hsu and P.W. Anderson. Phys. Rev. B {\bf 38} 745(1988).
\bibitem{su22} E. Dagotto, E. Fradkin, and A. Moreo. Phys. Rev. B {\bf 38} 2926(1988).
\bibitem{su23} G. Baskaran. \textit{Indian J. Phys.} {\bf 80} (6), 583-592(2006). 
\bibitem{bt}
G. Baskaran, E. Tosatti and L. Yu, Int. J. Mod. Phys. B, {\bf 2}, 555, 5(1988).
%\bibitem{senthil}
% Senthil. T, Fisher. M. P. A, Phys. Rev. B, {\bf 62}, 7850-7881(2000).
\bibitem{martson} J. B. Martson, Phys. Rev. Lett. {\bf 61}, 1914(1988).
\bibitem{fradkin}  E. Fradkin, {\it Field Theories of Condensed Matter Systems}, HarperCollins Canada / Perseus Books. 
\bibitem{nspin1} E. Lieb, F. Wu, Phys. Rev. Lett. {\bf 20}, 1444(1968).
\bibitem{nspin2} V. Kalmeyer, R. B. Laughlin, Phys. Rev. Lett. {\bf 59}, 2095(1987).
\bibitem{nspin3}P. W. Anderson, {\it Science} {\bf 235}, 1196(1987).
\bibitem{nspin4}P. W. Anderson, {\it Science} {\bf 288}, 480 (2000).
\bibitem{WenPSG}X. G. Wen, Phys. Rev. B {\bf 65}, 165113(2008).
\bibitem{lieb} E. H. Lieb, Phys. Rev. Lett. {\bf 73}, 2158 (1994).
\bibitem{wen-niu} X. G. Wen and Q. Niu, Phys. Rev. B {\bf 41}, 9377(1990).
\bibitem{abhinav} Abhinav Saket, S. R. Hassan and R. Shankar, 
 Phys. Rev. B {\bf  82}, 174409 (2010).
\bibitem{nayak} F. J. Burnell and Chetan Nayak, Phys. Rev. B {\bf 84}, 125125 (2011).
\end{thebibliography}
\end{document}